\documentclass[]{article}
\usepackage{graphicx}
\usepackage{subfig}
\usepackage{epstopdf}
\usepackage{epsfig}
\usepackage{authblk}
\usepackage[T1]{fontenc}
\usepackage[utf8]{inputenc}

\title{The Period Evolution of the Chemically Peculiar Star V473\,Tau}
\author[1,2]{D.Ozuyar\thanks{dozuyar@ankara.edu.tr}}
\author[2]{I. R. Stevens\thanks{irs@star.sr.bham.ac.uk}}

\affil[1]{Ankara University, Astronomy and Space Sci. Dept., Tandogan, Ankara, 06100, Turkey}
\affil[2]{The University of Birmingham, School of Physics and Astronomy, Birmingham, B15 2TT, UK}

\begin{document}

\maketitle

\begin{abstract}
In this paper, the period evolution of the rotating chemically peculiar star V473\,Tau (A0Si, V = 7.26 mag) is investigated. Even though the star has been observed for more than fifty years, for the first time four consecutive years of space-based data covering between 2007 and 2010 is presented. The data is from the {\sl STEREO} satellite, and is combined with the archival results. The analysis shows that the rotation period of V473\,Tau is $1.406829(10)$ days, and has slightly decreased with the variation rate of $-0.11(3)$~s~yr$^{-1}$ over time. Also, the acceleration timescale of the star is found to be around $-1.11(63) \times 10^6$~yr, shorter than its main sequence lifetime ($9.26(1.25) \times 10^8$~yr). This indicates that the process of decrease in period might be reversible. On this basis, it can be suggested that V473\,Tau has a possible magnetic breaking and a differential rotation, which cause a variation in the movement of inertia, and hence the observed period change.
\end{abstract}

\section{Introduction}
Chemically peculiar (CP) variables are spread between late-B and early-F spectral types, and thus contain various stars with effective temperatures greater than 6,500 K \cite{bib1}. These variables are comprised mostly of Ap and Bp stars, which differ from other types having the same temperature by their abnormal chemical compositions and slow rotations. The reason for the peculiarity is an under-abundance of solar-like elements, as well as an overabundance of both metal and rare-earth elements across their surfaces \cite{bib2}. Magnetic fields, radiative acceleration, and atomic diffusion determine the surface distribution of elements \cite{bib3}, and lead them to be present in the form of spots and rings on the surface. Along with rotation, these non-uniformly distributed regions cause periodic variations in magnetic fields, line profile, and energy distribution, as well as in photometric brightness (oblique-rotator model). The periods of these variations are generally between a day and a week. Depending on the slow rotation, surface spot regions can remain stable for decades. Such a situation enables remarkably accurate calculations of surface distribution, rotation period, and rotational breaking mechanisms. However, only very few of the CP stars discovered in our galaxy and others exhibit periodic variations, and less than one-tenth of these have been observed for scientific investigation. In order to study this type of stars, accurate observations are needed (accuracy $>$ 0.005 mag \cite{bib2}). The high-precision instruments of the {\sl STEREO} satellite are a quite suitable, space-based source, since seasonal and four-year {\sl STEREO} observations provide a precision of $2.0 \times10^{-4}$ and $7.0 \times 10^{-5}$ mmag, respectively.

\section{Literature Review}

The photometric variability of V473\,Tau (A0Si, V = 7.26 mag) was first detected by Burke et al. \cite{bib4}. They calculated the period of this variation as around 1.39(2) days, but this period value produced a light curve (LC) with a scattered maximum. Hence, Rakosch and Fiedler \cite{bib5} noted that their observations were more adaptable with a double period. Subsequently, Maitzen \cite{bib6} derived a rotation period of 2.7795(1) days, which was indeed twice that of previous values. As a result of the double period, two minima and maxima having different levels were formed in the LC; this situation was explained in terms of the different chemical regions on the surface. Most importantly, this was a significant case since a double wave structure was not a common condition among Si stars. In a recent study, Jerzykiewicz \cite{bib7} investigated rotation periods and found a value of 1.4068541(29) days in U, B, and V bands. However, he could not completely determine the origin of the variabilities as he was unable to conclude whether the star was an oblique rotator or a g-mode pulsator.

\section{STEREO Data and Data Analysis}

The data were provided from the HI-1A instrument on-board the {\sl STEREO-A} satellite. The HI-1A is capable of observing background stars with the magnitude of $12^m$ or brighter for a maximum of 20 days and a useful stellar photometer which covers the region around the ecliptic (20\% of the sky) with the field of view of $20^{\circ} \times 20^{\circ}$. The nominal exposure time of the camera is 40 seconds, and putting 30 exposures together on board, a 40-minute integrated cadence has been obtained to transmit for each HI-1 image \cite{bib8}. Therefore, the Nyquist frequency of the data is around 18~c~d$^{-1}$. LCs mostly affected by solar activities were cleaned with a $3^{rd}$ order polynomial fit. Observation points greater than $3 \sigma$ were clipped with a pipeline written in the Interactive Data Language (IDL) (For a more detailed description of the data preparation, refer to Sangaralingam \& Stevens \cite{bib9} and Whittaket et al. \cite{bib10}). The LC of V473\,Tau presented a sinusoidal characteristic due to spot modulation on the stellar surface. Therefore, all analyses were performed using the Lomb-Scargle (LS) algorithm since it is more sensitive to such variations. To determine a model of the sinusoidal LCs, the Levenberg-Marquardt Optimization method was applied, and the best fit was obtained after 5000 iterations. After the derivation of the model LC, random Gaussian noise with the mean of zero and the sigma value, which was determined from the cleaned curve, was produced and added to the model. This process was repeated 500 times. The most accurate frequencies and their uncertainties were assessed using the Monte-Carlo simulation algorithm. The results were compared to those derived from the Phase Dispersion Minimization (PDM) method and Period04. To perform O-C calculations and to investigate period variabilities over years, the best extremum times  were obtained from the seasonal LCs, and were put together with data from the literature.

\begin{table}[]
\small
\begin{center}
\caption{Frequency analysis results of V473\,Tau.}
\smallskip
\begin{tabular}{l l l l r} 
\hline
\noalign{\smallskip}
\textbf{V473\,Tau}	&\textbf{LS}&\textbf{Period04}&\textbf{PDM}&\textbf{Amp.}\\
&\textbf{(c~d$^{-1}$)}&\textbf{(c~d$^{-1}$)}&\textbf{(c~d$^{-1}$)}&\textbf{(mmag)}\\
\noalign{\smallskip}
\hline
\noalign{\smallskip}																	
2007	&	0.7104(8)	&	0.7101(8)	&	0.7120(16)	&	8.63(25)	\\ \noalign{\smallskip}
2008	&	0.7101(6)	&	0.7101(6)	&	0.7118(15)	&	10.93(24)	\\ \noalign{\smallskip}
2009	&	0.7116(7)	&	0.7116(8)	&	0.7157(15)	&	8.62(24)	\\ \noalign{\smallskip}
2010	&	0.7128(7)	&	0.7128(8)	&	0.7123(20)	&	9.20(25)      \\ \noalign{\smallskip}
Comb.	&	0.710818(5)	&	0.710818(7)	&	0.711164(5)	&	9.33(13)	\\ \noalign{\smallskip}
\hline 
\end{tabular}
\label{tab:table1}
\end{center}
\end{table}

\begin{figure}[!t]
\includegraphics[width =1.\textwidth]{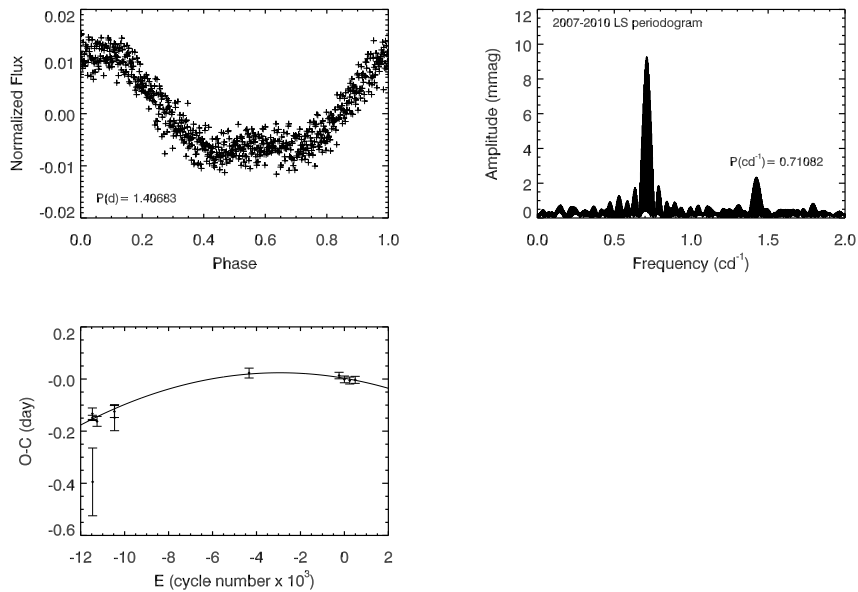}
\caption{Folded light curve produced by the LS period of 0.710818 c~d$^{-1}$ (top right), LS analysis of the four-year combined data (top left) and the O-C variation graphic of V473\,Tau (bottom).}
\label{fig:figure1}
\end{figure}

\section{Results and Discussion}

In this research, we obtained four consecutive years of data between 2007 and 2010. As reported by other researchers, all the LCs had explicit periodicity. Individual LS, PDM and Period04 analyses of the seasonal curves showed a frequency at around 0.71 c~d$^{-1}$ (Table~\ref{tab:table1}). Furthermore, we detected the existence of another strong peak at approximately 1.40 c~d$^{-1}$ (0.71 days) on the LS periodogram (Figure~\ref{fig:figure1}, top right).

Combining the four-year data, the precise rotation period of the star (0.710818 c~d$^{-1}$) was determined with the help of the LS method, and the main LC was plotted based on this value (Figure~\ref{fig:figure1}, top left). Accordingly, the folded LC was clearly formed by a maximum and a broad minimum. The maximum was quite strong and had a peaked top. Moreover, there was a barely detectable bump in the middle of the minimum, which is probably the cause of the peak at 1.40 c~d$^{-1}$ on the LS periodogram. Also, we produced a folded LC using the double {\sl STEREO} period since Maitzen \cite{bib6} noted that his observations were compatible with the period value of 2.7795 days. We derived a relatively clean LC with two minima and maxima. However, since the consecutive structures appeared similar to each other, we assumed that the period value of 0.710818 c~d$^{-1}$ was the full rotation period.

To investigate a possible period variation, the best maximum times of the seasonal LCs were derived (HJD24554583.4049, 24554922.4465, 24555274.1565, 24554241.5599), and these values were combined with the epochs from the literature. The archival epochs given as JD were converted to HJD. Based on Figure~\ref{fig:figure1} (bottom), we found out that the O-C diagram of the star exhibited a period decrease with the variation rate of around $-1.27(30) \times 10^{-6}$ d~y$^{-1}$ or $-0.11(3)$ s~y$^{-1}$ (parabolic line). With the help of the LS period and the {\sl STEREO} maximum time having the smallest error value, we determined the light elements as:   

\begin{equation}
\nonumber \\
HJD_{max} = 2454583.4049(129)+ 1.406829(10) E - 2.44(58) \times 10^{-9} E^2 ~.
\end{equation}

Since this star was a single rotating variable, such a period decrease might most likely be explained by an acceleration in rotation after a magnetic braking. Using the period and its variation rate, the acceleration time-scale of the star ($\tau_{ACC} = P/\dot{P}$) was determined as $\tau_{ACC} = -1.11(63) \times 10^{6}$ yr. We also found the main sequence lifetime of the star ($\tau_{MS} = \tau_{\odot}~(M/M_{\odot})^{-2.5}$) as $\tau_{MS} = 9.26(1.25) \times 10^{8}$ yr (Table~2).

\begin{table}
\small
\begin{center}  
\caption{Period and its variation rate as well as the acceleration and main sequence lifetime of V473\,Tau are presented in the upper raws. Physical parameters are given in the lower raws.}
\smallskip
\begin{tabular}{c c c c c c} \hline
\noalign{\smallskip}
$P$	&	$dP/dt$	&	$\dot{P}/P$	&	$\tau_{ACC}$	&	$\tau_{MS}$\\
 (day)	&	(s~yr$^{-1}$)	&	(s$^{-1}$)	&	(yr)	&	(yr)	\\
\noalign{\smallskip}
\hline
\noalign{\smallskip}

1.406830(10)	&	-0.11(3)	&	-2.86(68)$\times 10^{-14}$	&	-1.11(63)$\times 10^{6}$	&9.26(1.25)$\times 10^{8}$	\\
\noalign{\smallskip}
\hline\hline
\noalign{\smallskip}
\noalign{\smallskip}
$Log (L/$L$_{\odot})$	&	$Log (T)$	&	$M$	&	$R$	&	$V_{eq}$ \\
	&		&	(M$_{\odot}$)	&	(R$_{\odot}$)	&	(km~s$^{-1}$)	\\\hline
\noalign{\smallskip}
1.64(15)	&	4.045(11)	&	2.59(14)	&	1.80(32)	&	65(12)	\\
\hline

\end{tabular}
\end{center}
**Temperature, luminosity and mass values were adopted from Wraight et al. \protect\cite{bib14}. Radius and rotational velocities were estimated from the LS period, temperature, and luminosity values.
\label{tab:table2}
\end{table}

The variation in period of V473\,Tau ($\dot{P}/P$ in Table~2) is 10 times greater than that of the most massive CP stars given by Mikulasek et al. \cite{bib11}. In addition, its acceleration is nearly three orders of magnitude shorter than the main sequence lifetime. This, in turn, suggests that process of decrease in the period may be reversible. If so, the length of the cycle is roughly calculated as 92(11) yr (estimated by $T_{cyc} \sim P~\sqrt{2/\dot{P}}$; \cite{bib12}). Considering the fact that period variation processes may be reversible due to shorter acceleration time-scale than that of the main sequence lifetime, the rigid rotation hypothesis should be discarded and the differential rotation model should alternatively be discussed as expressed by Stepien \cite{bib13}. If a differential rotation is indeed in question for this star, the acceleration in the rotation may be interpreted as a consequence of torsional oscillations produced by meridional circulations being in interaction with a magnetic field, and of rotational braking in outer layers caused by angular momentum loss via magnetically-confined stellar wind \cite{bib13}.

\section*{Acknowledgments}
We acknowledge assistance from Vino Sangaralingam and Gemma Whittaker in the production of the data used in this study. The STEREO Heliospheric imager was developed by a collaboration that included the Rutherford Appleton Laboratory and the University of Birmingham, both in the United Kingdom, and the Centre Spatial de Lige (CSL), Belgium, and the US Naval Research Laboratory (NRL),Washington DC, USA. The STEREO/SECCHI project is an international consortium of the Naval Research Laboratory (USA), Lockheed Martin Solar and Astrophysics Lab (USA), NASA Goddard  Space Flight Center (USA), Rutherford Appleton Laboratory (UK), University of Birmingham (UK), Max-Planck-Institut fr Sonnen-systemforschung (Germany), Centre Spatial de Lige (Belgium), Institut dOptique Thorique et Applique (France) and Institut dAstrophysique Spatiale (France).

\end{document}